
\magnification = 1200
\def\lapp{\hbox{$ {
\lower.40ex\hbox{$<$}
\atop \raise.20ex\hbox{$\sim$}
}
$}  }
\def\rapp{\hbox{$ {
\lower.40ex\hbox{$>$}
\atop \raise.20ex\hbox{$\sim$}
}
$}  }
\def\barre#1{{\not\mathrel #1}}
\def\krig#1{\vbox{\ialign{\hfil##\hfil\crcr
$\raise0.3pt\hbox{$\scriptstyle \circ$}$\crcr\noalign
{\kern-0.02pt\nointerlineskip}
$\displaystyle{#1}$\crcr}}}
\def\upar#1{\vbox{\ialign{\hfil##\hfil\crcr
$\raise0.3pt\hbox{$\scriptstyle \leftrightarrow$}$\crcr\noalign
{\kern-0.02pt\nointerlineskip}
$\displaystyle{#1}$\crcr}}}
\def\ular#1{\vbox{\ialign{\hfil##\hfil\crcr
$\raise0.3pt\hbox{$\scriptstyle \leftarrow$}$\crcr\noalign
{\kern-0.02pt\nointerlineskip}
$\displaystyle{#1}$\crcr}}}

\def\svec#1{\skew{-2}\vec#1}
\def\Tr{\,{\rm Tr }\,}

\def\g5{\gamma_5}

\def\lp1{{\cal L}_{\pi N}^{(1)}}
\def\lp2{{\cal L}_{\pi N}^{(2)}}
\def\lp3{{\cal L}_{\pi N}^{(3)}}

\topskip=0.60truein
\leftskip=0.18truein
\vsize=8.8truein
\hsize=6.5truein
\tolerance 10000
\hfuzz=20pt

\baselineskip 14pt plus 1pt minus 1pt
\pageno=0
\centerline{\bf TOPICS IN CHIRAL PERTURBATION THEORY}
\vskip 48pt
\centerline{Ulf-G. Mei{\ss}ner}
\vskip 16pt
\centerline{{\it Centre de Recherches Nucl\'{e}aires et Universit\'{e}
Louis Pasteur de Strasbourg}}
\centerline{\it Physique Th\'{e}orique,
BP 20Cr, 67037 Strasbourg Cedex 2, France}
\vskip 1.0in
\centerline{ABSTRACT}
\medskip
\noindent I consider some selected topics in chiral perturbation theory (CHPT).
 For
the meson sector, emphasis is put on processes involving pions in the isospin
zero S-wave which require multi-loop calculations. The advantages and
shortcomings of heavy baryon CHPT are discussed. Some recent results on the
structure of the baryons are also presented.
\medskip
\vskip 1.0in
\centerline{REVIEW TALK}
\centerline{THIRD WORKSHOP ON HIGH ENERGY PARTICLE PHYSICS}
\centerline{Madras, India, January 1994}
\vfill
\noindent CRN--94/04  \hfill February 1994
\vskip 12pt
\eject
\baselineskip 14pt plus 1pt minus 1pt
\noindent{\bf 1. INTRODUCTION}
\bigskip
This talk will be concerned with certain aspects of the standard model in
the long--distance regime. I will argue that there exists a rigorous
calculational scheme and that plenty of interesting and {\it fundamental}
problems await a solution. I hope this will trigger further detailed
studies of these topics (and others which can only be mentioned en passant).

Our starting point is the observation that in
the three flavor sector, the QCD Hamiltonian can be written as
$$\eqalign{ H_{\rm QCD} &= H_{\rm QCD}^0 + H_{\rm QCD}^I \cr
H_{\rm QCD}^I &= \int d^3x \lbrace m_u \bar u u + m_d \bar d d + m_s \bar s s
\rbrace
\cr} \eqno(1)$$
with $H_{\rm QCD}^0$ symmetric under chiral SU(3)$_L \times$ SU(3)$_R$.
On a typical hadronic scale, say $M_\rho = 770$ MeV, the current
quark masses $m_q = m_u, m_d, m_s$ can be considered as perturbations. The
chiral symmetry of the Hamiltonian is spontaneously broken down to its
vectorial subgroup SU(3)$_V$ with the occurence of eight (almost) massless
pseudoscalar mesons, the Goldstone bosons
($\varphi = \pi^+,\pi^0,\pi^-,K^+,K^-,K^0,\bar{K}^0,\eta$)
$$ M^2_\varphi = m_q \, B + {\cal O}(m_q^2)  \eqno(2)$$
with $B = -<0|\bar q q|0>/F_\pi^2$ and $F_\pi$ the pion decay constant. In the
confinement (long-distance) regime, the properties of the standard model
related to this symmetry can be unambigously worked out in terms of an
effective Lagrangian,
$${\cal L}_{\rm QCD} = {\cal L}_{\rm eff} [U, \partial_\mu U, \ldots, {\cal M}]
\eqno(3)$$
with ${\cal M} = {\rm diag}(m_u,m_d,m_s)$ the quark mass matrix and the
Goldstone bosons are collected in the matrix-valued field $U(x) = \exp \lbrace
i \sum_{a=1}^8 \varphi_a (x) \lambda^a / F_\pi \rbrace $.
 Of course, there is an infinity of
possibilities of representing the non-linearly realized chiral symmetry. While
the QCD Lagrangian is formulated in terms of quark and gluon fields and the
rapid rise of the strong coupling constant $a_S (Q^2)$ with decreasing $Q^2$
forbids a systematic perturbative expansion, matters are different for the
effective field theory (EFT) based on the effective Lagrangian (3). It can be
written as a string of terms with increasing dimension,
$${\cal L}_{\rm eff} = {\cal L}_{\rm eff}^{(2)} +
{\cal L}_{\rm eff}^{(4)} + {\cal L}_{\rm eff}^{(6)} + \ldots \eqno(4) $$
if one counts the quark masses as energy squared. To lowest order, the
effective Lagrangian contains two parameters, $F_\pi$ and $B$. It is worth to
stress that $B$ never appears alone but only in combination with the quark mass
matrix, alas the pseudoscalar meson masses. Consequently, any matrix--element
$<ME>$ for the interactions between the pseudoscalars can be written as
$$<ME> = c_0 ({E \over \Lambda})^2 +
\biggl[ \sum_{i=1}^n (c_{1i}) + ({\rm non-local}) \biggr]
({E \over \Lambda})^4 + {\cal O}({E \over \Lambda})^6 \eqno(5)$$
This is obviously an energy expansion or, more precisely, a simultaneous
expansion in small external momenta {\it and} quark masses.
The first term on the r.h.s. of (5) leads to  nothing but the well--known
current algebra results, the pertinent coefficient $c_0$ can be entirely
expressed in terms of $F_\pi$, the Goldstone masses and some numerical
constants. As one of the most famous examples I quote Weinberg's result for
the S-wave, isospin zero scattering length [1],
$$ a_0^0 = {7 M_\pi^2 \over 32 \pi F_\pi^2} \eqno(6)$$
which is such an interesting observable
because it vanishes in the chiral limit $M_\pi \to 0$.
At next-to-leading order, life is somewhat more complicated. As first shown by
Weinberg [2] and discussed in detail by Gasser and Leutwyler [3], one has to
account for meson loops  which are naturally generated by the interactions.
These lead to what I called "non--local" in (5).  In
fact, it can be shown straightforwardly that any N--loop contribution is
suppressed with respect to the leading order result by $(E / \Lambda)^{2N}$.
At ${\cal O}(E^4)$, the loop contributions do not introduce
any new parameters. However, one also
has to account for the contact terms of dimension  four which are
accompanied by a priori unknown coupling constants (the $c_{1i}$ in (5)). These
so--called low--energy constants  serve to renormalize the infinities
related to the pion loops.
Their finite pieces are then fixed from some experimental input. In the case of
flavor SU(2), one has $n = 7$. Two of these constants are related to
interactions between the pseudoscalars, three to quark mass insertions and the
remaining two have to be determined from current matrix elements.
The inclusion of gauge boson couplings to the Goldstone
bosons is most simply and
economically done in the framework of external background sources. Notice also
that at order $E^4$ the chiral anomaly can be unambigously included in the
EFT. At order $E^6$, one has to consider loop diagrams with insertions from
${\cal L}_{\rm eff}^{(2)}$ and
${\cal L}_{\rm eff}^{(4)}$ as well as contact terms from
${\cal L}_{\rm eff}^{(6)}$ which introduces new couplings.
Once the low--energy constants are fixed, the aspects of the
dynamics of the standard model related to the chiral symmetry can be worked
out {\it systematically} and {\it unambigously}. Clearly, the EFT can only be
applied below a typical scale $\Lambda \simeq M_\rho$ and higher loop
calculations become more and more cumbersome (but can't always be avoided as
will be discussed below). This is the basic framework of CHPT in a nutshell.
For more details, I refer to refs.[2,3], my review [4] and the extensive list
of references given therein. It is worth pointing out that Leutwyler has
recently given a more sound foundation of the effective Lagrangian approach
by relating it directly to the pertinent  Ward--Identities [5].
\bigskip
\noindent{\bf 2. MESON--MESON SCATTERING AND THE MODE OF QUARK CONDENSATION}
\bigskip
Pion--pion and pion--kaon scattering are the purest reactions between the
pseudoscalar Goldstone bosons. The Goldstone theorem mandates that as the
energy goes to zero, the interaction between the pseudoscalars vanishes.
Consequently, $\pi \pi$ and $\pi K$ scattering are the optimal testing grounds
for CHPT.

Let me first consider the chiral expansion of the isospin zero
S--wave in $\pi \pi$ scattering. In the standard formulation of CHPT, Gasser
and Leutwyler have derived a low--energy theorem generalizing Weinberg's result
(6) [6],
$$a_0^0 = 0.20 \pm 0.01 \eqno(7)$$
which is compatible with the data, $a^0_0 = 0.23 \pm 0.08$ [7]. The theoretical
value (7) rests on the assumption that $B$ is large, i.e. of the order of 1 GeV
(from current values of the scalar quark condensates). However, if $B$ happens
to be small, say of the order of $F_\pi$, one has to generalize the CHPT
framework as proposed by Stern et al.[8]. In that case, the quark mass
expansion of the Goldstone bosons takes the form
$$ M^2_\varphi = m_q \, B + m_q^2 \, A + {\cal O}(m_q^3)  \eqno(8)$$
with the second term of comparable size to the first one.
In ref.[9], this framework is discussed in more detail and a novel
representation of the $\pi \pi$ amplitude which is exact including order $E^6$
and allows to represent the {\it whole} $\pi \pi$ scattering amplitude in terms
of the S-- and P--waves and six subtraction constants is given.
The presently available
data are not sufficiently accurate to disentangle these two possibilities. More
light might be shed on this when the $\phi$--factory DA$\Phi$NE
will be in operation
(via precise measurements of $K_{\ell 4}$--decays) or if the proposed
experiment
to measure the lifetime of pionic molecules [10] will be done. It should also
be pointed out that recent lattice QCD results seem to be at variance with the
expansion (8), but this can only be considered as indicative [11].
Also, the experimentally well--fulfilled GMO relation for the pseudoscalar
meson masses arises naturally in the conventional CHPT framework but requires
parameter fine--tuning in case of a small value of $B \sim 100$ MeV.
Novel high
precision experiments at {\it low} {\it energies} are called for. This is an
important question concerning our understanding of the standard model and it
definitively should deserve more attention. For more details, I refer to
sections 4.1 and 4.2 of ref.[4] as well as ref.[9].

In fig.1, I show the phase--shift $\delta^0_0$ from threshold (280 MeV) to
approximately 600 MeV [12]. One notices the rapid rise of the
phase shift, and at 600 MeV it is already as large as 55 degrees and passes
through 90 degrees at about 850 MeV. At energies below 600 MeV, the other
partial waves do not exceed 15 degrees (in magnitude). This behaviour
of $\delta^0_0$ is attributed to the so--called
strong pionic final state interactions which I will discuss in section 3.

\midinsert
\vskip 10.0truecm
{\noindent\narrower \it Fig.~1:\quad
$\pi \pi$ scattering phase shift $\delta_0^0 (s)$.
The dashed line gives the tree result and the dashed--dotted  the
one--loop prediction. Also shown is the Roy equation band.
The data can be traced back from ref.[12].
The double--dashed line corresponds to the one--loop result
based on another definition of the phase--shift which differs at order $E^6$
from the one leading to the dashed--dotted line (and thus gives a measure of
higher order corrections). On the right side of the hatched area, the one--loop
corrections exceed 50 per cent of the tree result.
\smallskip}
\endinsert

\noindent As indicated in fig.1,
beyond 450 MeV the one loop corrections are half as big
as the tree phase. Nevertheless, one can make a rather precise statement about
the phase of the CP--violation parameter $\epsilon'$[12],
$$\Phi(\epsilon') = {\pi \over 2} -(\delta^0_0 - \delta_0^2)\biggr|_{s =
M_{K^0}^2} = (45 \pm 6)^\circ  \eqno(9)$$
This is due to the fact that the corrections to $\delta_0^2$ are of the same
sign as the ones to $\delta^0_0$ and thus cancel. At tree level, $\Phi
(\epsilon') = 37^\circ$. The accuracy of the theoretical
prediction is as good as the
resent empirical one, $\Phi (\epsilon')_{\rm exp} = (43 \pm 8)^\circ$ [7].
Notice that it is much more difficult to get a precise number on $\Phi
(\epsilon')$ from $K \to 2 \pi$ decays because of the variety of isospin
breaking effects one has to account for (this theme is touched upon in
ref.[12]).

I briefly turn to the case of $\pi K$ scattering. Here, the empirical
situation is even worse, which is very unfortunate. In the framework of
conventional CHPT, the threshold behaviour of the low partial waves can be
unambigously predicted [13] since all low--energy constants in SU(3) are fixed.
Furthermore, since the mass of the strange quark is of the order of the QCD
scale--parameter, it is less obvious that the chiral expansion at
next--to--leading order will be sufficiently accurate.
Much improved empirical
information of these threshold parameters might therefore lead to a better
understanding of the three flavor CHPT. Another possibility is that the
threshold of $\pi K$ scattering at 635 MeV is alreday so high that one has
to connect CHPT constraints with dispersion theory. This concept has
investigated in detail by Dobado and Pelaez [14] and certainly improves the
prediction in the P--wave drastically. Another way of extending the EFT through
the implicit inclusion of resonance degrees of freedom is discussed in
ref.[62]. On the experimental side, a measurement of $\pi K$ molecule decays
would certainly help to clarify the situation [11].
\bigskip
\noindent{\bf 3. TWO LOOPS AND BEYOND}
\bigskip
The simplest object to study in detail the strong pionic final state
interactions in the isospin zero S--wave is a three--point function, namely
the so--called scalar form factor (ff) of the pion,
$$ <\pi^a (p') \pi^b (p) | \hat{m} (\bar u u + \bar d d) |0> = \delta^{ab} \,
\Gamma_\pi (s) M_\pi^2    \eqno(10)$$
with $s = (p'+p)^2$. To one loop order, the scalar ff
$\Gamma_{\pi,2} (s)$ has been given in
ref.[3]. As shown in fig.2, closely about the two--pion cut, the real as well
as the imaginary part of the one loop representation are at variance with the
empirical information obtained from a dispersion--theoretical analysis [15].
However, unitarity allows one to write down a two--loop representation [16],
$$\Gamma_\pi (s) = d_0 + d_1 \, s + d_2 \, s^2 + {s^3 \over \pi}
\int_{4M_\pi^2}^\infty {ds' \over s'^3} {\sigma (s') \over s' -s }
\biggl\lbrace
T^0_{0,2} ( 1 + {\rm Re}\Gamma_{\pi,2} ) + T^0_{0,4} \biggr\rbrace \eqno(11) $$
where $T^0_{0,2}$ and $T^0_{0,4}$ are the tree and one loop representations of
the $\pi \pi$ S--wave, isospin zero scattering matrix. Notice that the
imaginary part of $\Gamma_\pi (s)$ to two loops is entirely given in terms of
known one loop amplitudes. The three subtraction constants appearing
in (11) can be fixed from the empirical knowledge of the normalization, the
slope and the curvature of the scalar ff at the origin. In the chiral
expansion, these numbers are combinations of two low--energy constants from
${\cal L}_{\rm eff}^{(4)}$ and two from
${\cal L}_{\rm eff}^{(6)}$.

\midinsert
\vskip 11.0truecm
{\noindent\narrower \it Fig.~2:\quad
Scalar form factor of the pion. The curves labelled
'1', '2', 'O' and 'B' correspond to the chiral prediction to one--loop,
to two--loops, the modified Omn\`es representation and the result of the
dispersive analysis, respectively. The real part is shown in (a) and
the imaginary part in (b).
\smallskip}
\endinsert

The turnover of the scalar ff at around 550 MeV can
be understood if one rewrites (11) in an exponential form,
$$ {\rm Re} \Gamma_\pi (s) = P(s) \exp [{\rm Re}\Delta_0 (s)] \cos\delta^0_0
+ {\cal O}(E^6)   \eqno(12)$$
with Im $\Delta_0 (s) = \delta^0_0 + {\cal O}(E^6)$ fulfilling the final--state
theorem at next--to--leading order. Although this representation is not unique,
it allows to understand the vanishing of Re$\Gamma_\pi (s)$ at 680 MeV since
the phase (in the loop approximation) passes through $90^\circ$ at this energy
thus forcing the turnover. Expanding $\cos \delta^0_0 = 1 - (\delta^0_0)^2 +
\ldots = 1 + {\cal O}(s^2 / F_\pi^4)$ it becomes clear why this behaviour can
only show up at two loop order (and higher). One can do even better and sum up
all leading and next--to--leading logarithms by means of an Omn{\`e}s
representation [16]. This leads to a further improvement in Re$\Gamma_\pi (s)$
and allows to understand that the very accurate two loop result for
Im$\Gamma_\pi (s)$ is not spoiled by higher orders, these can be estimated from
the improved chiral expansion of the scalar ff and are found to be small below
550 MeV. The physics behind all this is that the two--loop corrections lead to
the two--pion cut with proper strength which dominates the scalar ff below 600
MeV. To go further one would have to include inelasticities (which start at
order $E^8$), in particular the strong coupling to the $\bar K K$ channel. It
is also worth pointing out that the scalar ff can only be represented by a
polynomial below $s = 4 M_\pi^2$. Notice that in this energy range the
normalized scalar ff varies from 1 to 1.4, signaling a large scalar radius of
the pion. For comparison, the vector ff  changes from 1 to 1.15 for $0 \le s
\le 4M_\pi^2$. In this way, unitarity allows to extend the  range of CHPT,
however, one has to be able to fix the pertinent subtraction constants (which
is the equivalent to determining the corresponding low--energy constants).

Another reaction which has attracted much attention recently is $\gamma \gamma
\to \pi^0 \pi^0$ in the threshold region. It belongs to the rare class of
processes which are vanishing at tree level (since the photon can only couple
to charged pions, one needs at least one loop) and do not involve any of the
low--energy couplings from ${\cal L}_{\rm eff}^{(4)}$ at one loop order.
 Some years ago, Bijnens and
Cornet [17] and Donoghue, Holstein and Lin [18] calculated the one--loop cross
section and found that it is at variance with the Crystal ball data [19] even
close to threshold (see fig.3).

\midinsert
\vskip 10.0truecm
{\noindent\narrower \it Fig.~3:\quad
Cross section for $\gamma \gamma \to \pi^0 \pi^0$. The chiral one and two loop
predictions are given by the dotted and the solid line, in order. The hashed
area is a dispersion--theoretical fit. The Crystal ball data are also shown.
{}From [22].\smallskip}
\vskip -0.5truecm
\endinsert

This is another case where one has to account for the
strong pionic final state interactions. At 400 MeV, one has
$$\biggl({\sigma^{\exp} \over \sigma^{\rm 1-loop}}\biggr)^{1/2}
= 1.3 \eqno(13)$$
which is a typical  correction in this channel (see discussion above
on $a_0^0$ and the scalar ff).
In fact, dispersion theoretical
calculations supplemented with current algebra constraints by Pennington [20]
tend to give the trend of the data (see the shaded area in fig.3). An improved
combination of chiral machinery and dispersion theory has been given by
Donoghue and Holstein [21]. Even better, Bellucci, Gasser and Sainio [22] have
performed a full two loop calculation. It involves some massive algebra and
three new low--energy constants have been estimated from resonance exchange
(the main contribution comes form the $\omega$). These couplings play, however,
no role below 400 MeV. The solid line in fig.3 shows the two--loop result for
the central values of the coupling constants. One finds a good agreement
with the data up to $E_{\pi \pi} = 700$ MeV. This resolves the long--standing
discrepancy between the chiral prediction and the data in the threshold region.
For a more detailed discussion of these topics and the related neutral pion
polarizabilities, I refer to ref.[22].

The last topic I briefly want to mention is the radiative kaon decay $K_L \to
\pi^0 \gamma \gamma$ which has no tree--level contribution and is given by a
finite one--loop calculation at order $E^4$ [23]. The predicted two--photon
invariant mass spectrum turned out to be in amazing agreement with the later
measurements [24].
However, the branching ratio which is also predicted was found
 about a  factor three too
small. Again, unitarity corrections work in the right direction. In recent
work by Cohen, Ecker and Pich [25] and Kambor and Holstein [26] it is shown
that
unitarity corrections (eventually supplemented by a sizeable $E^6$ vector meson
exchange contribution)
can indeed close the gap between the empirical branching
ratio and the CHPT prediction though not completely. These calculations are,
however, not taking into account all effects beyond $E^4$ but they underline
the importance of making use of dispersion theory in connection with CHPT.
\bigskip
\noindent{\bf 4. INCLUSION OF BAYONS}  \bigskip
While the chiral Lagrangian is particularly suited to investigate the
properties of the pseudoscalar mesons, it can also be used to gain insight into
the structure of the low--lying baryons.  I will be brief on the formal
aspects but rather refer to the reviews [4,27] and the extensive papers by
Gasser, Sainio and $\check {\rm S}$varc [28] and Krause [29].

Let me first restrict myself to the two flavor sector, the pion--nucleon
system. To lowest order ${\cal O}(E)$, the effective Lagrangian takes the
form
$${\cal L}_{\pi N} = \bar{\Psi} \bigl( i \gamma_\mu D^\mu - m + {1 \over 2}
g_A \gamma_\mu \gamma_5 u^\mu \bigr) \Psi  \eqno(14) $$
with $D_\mu$ the covariant derivative, $m$ the nucleon mass (in the chiral
limit), $g_A$ the axial--vector coupling constant (in the chiral limit) and
$u_\mu = i u^\dagger \nabla_\mu U u^\dagger$, $u = \sqrt{U}$ and $\Psi$
embodies
the proton and neutron fields. The physics becomes most transparent if one
expands the various terms in powers of the pion and external fields (like e.g.
the photon). The vectorial coupling includes e.g. the photon--nucleon vertex
and the two--pion seagull ("Weinberg term") whereas the axial--vector term
in (14) leads to the pseudovector $\pi N$ coupling and the famous
Kroll--Rudermann vertex (among others). The presence of the nucleon mass term,
which is of comparable size to the chiral symmetry breaking scale, does not
allow the nucleon four--momentum to be treated as small. This spoils the
one--to--one correspondence between the loop and the energy expansion (for more
details see refs.[27,28]). As pointed out in particular by Jenkins and Manohar
[30], heavy quark EFT methods help to overcome this problem. Consider the
nucleon as a very heavy, static source, i.e. non--relativistically. In that
case, one can write its four--momentum as
$$p_\mu = m v_\mu + l_\mu  \eqno(15)$$
with $v_\mu$ the four--velocity ($v^2 = 1$) and $l_\mu$ a small off--shell
momemtum, $v \cdot l \ll m$. One can therefore write the nucleon wave function
in terms of velocity eigenstates,
$$\Psi = \exp \bigl\lbrace i m v \cdot x \bigr\rbrace (H + h)   \eqno(16)$$
with $\barre v H = H$ and $\barre v h = -h$ (notice that I have interchanged
the $H$ and the $h$ in comparison to the standard (funny) notation). If one
now eliminates $h$ by use of its equation of motion (or by a Foldy--Wouthuysen
transformation), one finds (for details, see e.g. ref.[31])
$${\cal L}_{\pi N} = \bar{H} \bigl( i v \cdot D +
g_A u \cdot S \bigr) H + {\cal O}(1/m)  \eqno(17) $$
with $S_\mu$ the covariant spin--operator, $S_\mu = i \gamma_5 \sigma_{\mu \nu}
v^\nu / 2$, known to our anchestors also as the Pauli--Lubanski vector.
The cumbersome baryon mass term has disappeared and thus a
consistent power counting emerges (this is discussed  very nicely in Ecker's
lectures [32]). One also notices that all Dirac bilinears can be expressed in
terms of $v_\mu$ and $S_\mu$ thus faciliating the algebra enormeously. The
one loop graphs contribute at order $E^3$. As will be discussed below, it is
however mandatory to include the terms of order $E^4$ in the effective
Lagrangian for accurate one--loop calculations, i.e.
$${\cal L}_{\pi N} = {\cal L}_{\pi N}^{(1)} + {\cal L}_{\pi N}^{(2)}
+ {\cal L}_{\pi N}^{(3)} + {\cal L}_{\pi N}^{(4)}         \eqno(18)$$
where I have not exhibited the meson Lagrangian discussed before. The
coefficients accompanying ${\cal L}_{\pi N}^{(2)}$ are all finite since the
loops start to contribute at order $E^3$. A complete analysis of the divergence
structure at order $E^3$ will soon be available [33]. A systematic analysis of
nucleon properties to order $E^3$ can be found in [31] and some $E^4$
calculations have recently been performed. I will discuss one particular
case in some more detail below. I will also elaborate on two yet unsolved
problems in the heavy mass approach, one is related to the analytical structure
of S--matrix elements (which does not appear in the relativistic formulation,
see section 8) and the
other is the extension to flavor SU(3) and the inclusion of decuplet
fields (see sections 6 and 7).
\bigskip
\noindent{\bf 5. BARYON COMPTON SCATTERING}
\bigskip
Compton scattering off the nucleon at low energies offers important information
about the structure of these particles in the non-perturbative regime of QCD.
The spin--averaged forward scattering amplitude for real photons in the nucleon
rest frame can be expanded as a power series in the photon energy $\omega$,
$$ T (\omega ) = f_1 (\omega^2 ) \,
\vec{\epsilon}_f^{\, *} \cdot \vec{\epsilon}_i ,  \quad
f_1 (\omega^2 ) = a_0 + a_1 \, \omega^2 + a_2 \, \omega^4 + \ldots
\eqno(19)$$
where $\vec{\epsilon}_{i,f}$ are the polarization vectors of the initial and
final photon, respectively, and due to crossing symmetry only even powers of
$\omega$ occur. The Taylor coefficients $a_i$ encode the
information about the nucleon structure. The first term in eq.(19), $a_0 = -e^2
Z^2 / 4 \pi m$, dominates as the photon energy  approaches zero, it is only
sensitive to the charge  $Z$ and the mass $m$ of the
particle the photon scatters off (the Thomson limit). The term quadratic in
the energy is equal to the sum of the so-called electric ($\bar\alpha$) and
magnetic ($\bar\beta$) Compton polarizabilities, $a_1 = \bar \alpha + \bar
\beta$.  Corrections of higher order in
$\omega$ start out with the term proportional to $a_2$.
Over the last years, high precision measurements at Mainz,
Illinois, Oak Ridge and Saskatoon [34] have lead to the following empirical
values:
$\bar \alpha_p
= (10.4 \pm 0.6)
\cdot 10^{-4} {\rm fm}^3 \, ,  \bar \beta_p   = (3.8 \mp 0.6)
\cdot 10^{-4} {\rm fm}^3 \, , \bar \alpha_n  = (12.3  \pm 1.3)
\cdot 10^{-4} {\rm fm}^3 \, , \bar \beta_n   = (3.5 \mp 1.3)
\cdot 10^{-4} {\rm fm}^3$
making use of    the dispersion sum rules [35]
$(\bar \alpha+ \bar \beta)_p = (14.2 \pm 0.3) \cdot 10^{-4}$ fm$^3$ and
$(\bar \alpha+ \bar \beta)_n = (15.8 \pm 0.5) \cdot 10^{-4}$
fm$^3$.  The two outstanding features of these numbers are the fact that
$(\bar \alpha+ \bar \beta)_p \simeq (\bar \alpha+ \bar \beta)_n$ and that the
proton as well as the neutron behave essentially as (induced) electric dipoles.

In CHPT, it was found that to leading order in the chiral expansion the nucleon
em polarizabilities are given by a few one loop diagrams
(whose sum is finite) [36] with no
counter term contributions (much like the reactions $K_L \to \pi^0 \gamma
\gamma$ and $\gamma \gamma \to \pi^0 \pi^0$ discussed before). This calculation
was later redone in the heavy mass approach [31]. The pertinent diagrams are
shown in fig.4.

\midinsert
\vskip 5.0truecm
{\noindent\narrower \it Fig.~4:\quad
One--loop diagrams which lead to the nucleom em polarizabilities (20).
\smallskip}
\endinsert

By isospin arguments, one finds that they will lead to the same
polarizabilities for the proton and the neutron. The resulting expressions for
$\bar{\alpha}_{p,n}$ and $\bar{\beta}_{p,n}$ contain therefore only parameters
from the lowest order effective Lagrangian ${\cal L}^{(1)}_{\pi N} + {\cal
L}^{(2)}_{\pi \pi}$ [31,36],
$$ \bar{\alpha}_p = \bar{\alpha}_n = {5 e^2 g_A^2 \over 384 \pi^2 F_\pi^2}{1
\over M_\pi} = 12.2 \cdot 10^{-4} {\rm fm}^3 , \quad \bar{\beta}_p =
\bar{\beta}_n = {\bar{\alpha}_p \over 10} = 1.2 \cdot 10^{-4} {\rm fm}^3
\eqno(20)$$
In the chiral limit, the em polarizabilities diverge. This is expected since
the Yukawa suppression for massive pions turns into a long--range power--law
fall--off as $M_\pi \to 0$. Clearly, the leading order CHPT results (20)
explain the trends of the data. However, one might argue that the result for
the
magnetic polarizabilities is not very meaningful since one has not accounted
for
the strong $N \Delta$ M1 transition. In fact, this starts to contribute at
order $E^4$ (and higher) in agreement with the decoupling theorem [37]. In
ref.[38], a {\it complete} calculation of the em polarizabilities
to ${\cal O}(E^4)$ was given. At
this order, contact terms from ${\cal L}_{\pi N}^{(2,3,4)}$ enter. Some of them
can be directly related to  empirical information (viz $\pi N$
scattering [39]), others are estimated from resonance exchange, and this is
where the $\Delta (1232)$ comes in. It should be stressed that the resonance
saturation hypothesis to understand the values of the low--energy constants has
only strictly been tested in the meson sector [40]. In the absence of
sufficiently many accurate low--energy data, it serves as a working hypothesis
in the baryon sector (this situation will be improved when more data will
become available). The em polarizabilities to order $E^4$ take the form
$$(\bar{\alpha} , \bar{\beta})_{p,n} = {C_1 \over M_\pi} + C_2 \, \ln M_\pi
+ C_3   \eqno(21)$$
with $C_1 = 5 e^2 g_A^2 / 384 \pi^2 F_\pi^2$. The coefficient $C_2$ contains
some low--energy constants from ${\cal L}_{\pi N}^{(2)}$ and $C_3$ four novel
ones from ${\cal L}_{\pi N}^{(4)}$.  The second and the third term in (21) are
the new $E^4$ contributions. The $\Delta(1232)$ strongly dominates the constant
$C_3$. The non--analytic loop contribution $\sim \ln M_\pi$ is potentially
large. Indeed, in the case of $\bar{\beta}_p$ the $\ln$ contribution is
negative with a large coefficient and cancels most of the large positive one
related to the $\Delta(1232)$ exchange. In ref.[41], a thorough study of the
theoretical uncertainties entering the $E^4$ calculation was performed and the
following results emerged
$$\bar \alpha_p
= 10.5 \pm 2.0 \, ,  \bar \beta_p   = 3.5 \pm 3.6
\, , \bar \alpha_n  = 13.4  \pm 1.5 \, , \bar \beta_n   = 7.8 \pm 3.6
 \, , \eqno(22)$$
all in $10^{-4} {\rm fm}^3$.
For the electric polarizabilities the chiral expansion is well--behaved, i.e.
the $E^4$ corrctions amount to 14 (10) per cent for the proton (neutron). In
the case of the magnetic polarizabilities large cancellations occur and a
calculation at ${\cal O}(E^5)$ is called for. The large theoretical
uncertainties stem mostly from the badly knwon off--shell parameters related to
the $\pi N \Delta$ and $\gamma N \Delta$ dynamics and to some extent also from
the contribution from strange ($K^+$) loops.

In fig.5, I show the real part of the proton forward Compton scattering
amplitude $A_p(\omega) = - 4 \pi \, f_1(\omega^2)$ in comparison to the data
which follow from the total photonucleon absorption cross section via
$$
{\rm Re}\,A_p(\omega) = {e^2 Z^2 \over m} - {2 \omega^2 \over \pi}\, {\cal P}
\int_{\omega_0}^\infty \, d\omega' {\sigma_{\rm tot}(\omega') \over {\omega'}^2
- \omega^2}  \eqno(23)$$

\midinsert
\vskip 17.0truecm
{\noindent\narrower \it Fig.~5:\quad
Forward spin--averaged Compton amplitude for the proton in comparison to the
data. (a) $0 \le \omega \le 140$ MeV and (b) $ 140 \le \omega \le 210$ MeV.
\smallskip}
\vskip -0.5truecm
\endinsert
The amplitude has a branch point related to the threshold energy for single
pion
photoproduction at
$$\omega_0 = M_\pi(1 + {M_\pi \over 2 m}) \,\, . \eqno(24)$$
Notice that to order $E^3$ the chiral representation [31] has a cut
starting at $\omega_0
= M_\pi$ and only at ${\cal O}(E^4)$ one recovers the recoil correction $\sim
1/m$ in (24). This problem is inherent to the heavy mass formulation of baryon
CHPT. In the relativistic formulation, the corresponding analytic structures
(location of cut singularities)
are always given correctly and do not depend on the order of the chiral
expansion. I will pick up this theme in section 8. Fig.5 also shows that the
$E^4$ result for $A_p (\omega)$ reproduces the cusp at $\omega_0$. However,
above that energy, the chiral prediction is at variance with the data. This can
be traced back to the fact that the imaginary part changes sign at $\omega
\simeq 180$ MeV. One would have to go to two loops to get an improved
prediction for the imaginary part as alreday stressed in section 3.

It is fairly straightforward to extend the lowest order $E^3$
calculation to the three
flavor sector. This allows to predict the hyperon polarizabilities which
eventually will be measured at Fermilab and CERN via the Primakoff effect.
In SU(3), one
has two axial couplings and thus the lowest order  effective
meson--baryon Lagrangian reads
$${\cal L}^{(1)}_{MB}= \Tr(\bar B \,i v\cdot{\cal D} \,B) + D\, \Tr (\bar B
S^\mu \{ u_\mu , B\} ) + F \, \Tr ( \bar B S^\mu [ u_\mu, B] ) \eqno(25)$$
where $U(x)$ now contains the eight pseudoscalar fields ($\pi, K, \eta$) and
$B$
is a $3 \times 3$ matrix containing the low--lying baryon octet,
$$ B =	\left(
\matrix  { {1\over \sqrt 2} \Sigma^0 + {1 \over \sqrt 6} \Lambda
&\Sigma^+ &  p \cr
\Sigma^-
    & -{1\over \sqrt 2} \Sigma^0 + {1 \over \sqrt 6} \Lambda & n \cr
\Xi^-
	&	\Xi^0 &- {2 \over \sqrt 6} \Lambda \cr} \right)
\eqno(26)$$
The numerical values for the two axial couplings $D$ and $F$ are $D \simeq 3/4$
and $F \simeq 1/2$ subject to the constraint $D+F = g_A = 1.26$. The
hyperon polarizabilities can be calculated from the diagrams in fig.4 with
$\pi$ and $K$ loops and one finds e.g. [42]
$$\bar{\alpha}_{\Sigma^-} = 6 , \quad  \bar{\alpha}_{\Sigma^+} = 9 ,
\eqno(27)$$
(in canonical units), i.e. the $\Sigma^+$ is expected to have a larger electric
polarizability then the $\Sigma^-$ due to the kaon loop contributions.
Quark model estimates give a similar pattern [43].
The $\Sigma^+$ is made of $u$ and
$s$ quarks (which have opposite charges) and this allows for
internal electric dipole
excitations. In contrast, the
charge--like $d$ and $s$ quarks in the $\Sigma^-$ tend to
hinder such excitations leading to a small electric polarizability. The numbers
given in (27) should be considered as first estimates since a complete $E^4$
calculation in SU(3) has yet to be performed. I will now take  a critical
look at the status of three flavor baryon CHPT.
\bigskip
\goodbreak
\noindent{\bf 6. BARYON MASSES AND $\sigma$--TERMS} \bigskip
The simplest observbles to investigate in flavor SU(3) are the baryon masses
$m_N, m_\Lambda, m_\Sigma, m_\Xi$ and the three
proton $\sigma$--terms defined via
$$\eqalign{
\sigma_{\pi N}(t) &= \hat m <p'|\bar u u + \bar d d| p>\cr
\sigma_{KN}^{(1)}(t)&={1\over 2}(\hat m + m_s) <p'|\bar u u + \bar s s | p> \cr
\sigma^{(2)}_{KN}(t) &=  {1\over 2}(\hat m + m_s) <p'|-\bar u u + 2 \bar d d +
\bar s s| p> \cr } \eqno(28)$$
with $t = (p'-p)^2$ the invariant momentum transfer squared
and $\hat m = (m_u +
m_d)/2$ the average light quark mass. At zero momentum transfer, the strange
quark contribution to the proton mass is given by [44]
$$m_s <p|\bar s s |p> = \biggl({1\over 2} - {M_\pi^2 \over 4M_K^2} \biggr)
\biggl[3\sigma^{(1)}_{KN}(0)+ \sigma_{KN}^{(2)}(0) \biggr]+ \biggl({1\over 2} -
{M_K^2 \over M_\pi^2 } \biggr) \sigma_{\pi N}(0) \eqno(29)$$
making use of the leading order meson mass formulae $M_\pi^2 = 2 \hat m B $
and $M_K^2 = (\hat m + m_s) B$. This defines the scalar sector of baryon CHPT.
To calculate the mass spectrum to order $E^3$, we need the symmetry breaking
terms from
$${\cal L}_{MB}^{(2)} = b_D\, \Tr (\bar B \{ \chi_+ , B \} ) + b_F \, \Tr
(\bar B [ \chi_+,B]) + b_0 \, \Tr (\bar BB ) \Tr (\chi_+) \eqno(30)$$
with $\chi_+ = u^\dagger \chi u^\dagger + u \chi^\dagger u$ and $\chi = 2 B (
{\cal M} + {\cal S})$ where $\cal S$ denotes the nonet of external scalar
sources.  The constants $b_D$, $b_F$ and $b_0$ can be
fixed from the knowledge of the baryon masses and the $\pi N$ $\sigma$-term (or
one of the $KN$ $\sigma$-terms). The constant $b_0$ can not be determined from
the baryon mass spectrum alone since it contributes to all octet members in the
same way. To this order in the chiral expansion, any baryon mass takes the
form [44,45]
$$m_B = m_0 - {1\over 24 \pi F_\pi^2}\bigl[ \alpha^\pi_B M_\pi^3 + \alpha^K_B
M_K^3 + \alpha^\eta_B M_\eta^3 \bigr] + \gamma^D_B b_D + \gamma^F_B b_F
-2b_0(M_\pi^2 + 2M_K^2) \eqno(31)$$
The first term on the right hand side of (31) is the average octet mass
in the chiral limit, the second one
comprises the Goldstone boson loop contributions and the third
term stems from the counter terms in (30) ("resonance physics").
Notice that the loop contribution is
ultraviolet finite and non-analytic in the quark masses since $M_\phi^3 \sim
m_q^{3/2}$. The constants $b_D$, $b_F$ and $b_0$ are therefore finite.
A typical result at ${\cal O}(E^3)$ from a least--square fit to $m_N$,
$m_\Lambda$, $m_\Sigma$, $m_\Xi$ and $\sigma_{\pi N} (0) = 45$ Mev [46] is [44]
$$\eqalign{
 m_N &= (0.965 - 0.018 - 0.264 + 0.248) \,{\rm GeV} = 0.936 \, {\rm
GeV}   \cr
 m_\Lambda &= (0.965 - 0.006 - 0.588 + 0.743) \,{\rm GeV} = 1.141 \, {\rm
GeV} \cr}                 \eqno(32)$$
where the various terms are the average octet mass, the pion loop, the $K$ and
$\eta$ loop and the counterterm contributions, in order.
A closer look at the results (32) reveals
that there are large cancellations between the strange loops and the counter
terms. To have a more well--behaved chiral expansion, one might want to include
the low--lying decuplet baryons as
will be discussed below.
At this order and within the accuracy of the $E^3$ calculation,
the $KN$ $\sigma$-terms turn out to be
$$\sigma^{(1)}_{KN}(0) \simeq 200 \pm 50 \,{\rm MeV} , \quad
\sigma^{(2)}_{KN}(0) \simeq 140 \pm 40 \,{\rm MeV} \eqno(34)$$
which is comparable to the first order perturbation theory analysis having no
strange quarks, $\sigma^{(1)}_{KN}(0) = 205$ MeV and $\sigma^{(2)}_{KN}(0) =
63$ MeV [47]. At present, the $KN$ $\sigma$--terms are not well determined.
Since most of the phase shift data stem from kaon--nucleus scattering, it
is of advantage to define them in terms of nuclear isospin,
 $\sigma'_{KN} = (3\sigma_{KN}^{(2)}+
\sigma_{KN}^{(1)}$)/4 and  $\sigma''_{KN} = (\sigma_{KN}^{(2)} -
\sigma_{KN}^{(1)})/2$. The best determinations available gives $\sigma'_{KN}
(0) = 599 \pm 377$ MeV and $\sigma''_{KN} (0) = 87 \pm 66$ MeV [48]. This
translates into $\sigma^{(1)}_{KN} (0)= 469 \pm 390$ MeV and
 $\sigma^{(2)}_{KN} (0)= 643 \pm 378$ MeV.
Let me finish this section with a remark on the calculation of
 $m_N$ in a recent lattice simulation in quenched
baryon CHPT [49] (which is discussed in some detail by Golterman in these
proceedings),
$$\eqalign{
 m_N^{\rm CHPT} &= (0.97 - 0.5 M_\pi + 3.4 M_\pi^2 - 1.5 M_\pi^3) \,{\rm
GeV} \cr
 m_N^{\rm LFIT} &= (0.96 - 1.0 M_\pi + 3.6 M_\pi^2 - 2.0 M_\pi^3) \,{\rm GeV}
 \cr}        \eqno(34)$$  where LFIT denotes the fit to the lattice data.
It is most significant that the negative curvature due to the $M_\pi^3$ term
from quenched CHPT and from the lattice fit are of the same magnitude. There
is, however, a problem at small $M_\pi$. The lattice data give a too large
pion mass so that one has to lot $m_N$ versus $M_\pi$ and interpolate to the
physical pion mass. On finds a hook at the lower end of this curve which sheds
some doubts on the accuracy of the recently reported results by Butler et al.
[63]. For more details on this, see Golterman [64].
\bigskip  \goodbreak
\noindent{\bf 7. INCLUSION OF THE DECUPLET IN THE EFT}
\bigskip
The low-lying decuplet is only separated by $\Delta = 231$ MeV from the octet
baryons, which is just $(5/2) F_\pi$ (notice that I have not used the
conventional argument $\Delta = (5/3)M_\pi$ since the splitting stays finite
in the chiral limit much like $F_\pi$ and not at all like $M_\pi$)
and considerably smaller than the
kaon or eta mass. One therefore expects the excitations of these resonances to
play an important role even at low energies. This is also backed by
phenomenological models of the nucleon in which the $\Delta(1232)$ excitations
play an important role. In addition, there are large $N_C$ arguments [50]
which,
however, have to be taken {\it cum} {\it grano} {\it salis} since the chiral
and infinite number of colors limites do not commute.
 In the meson sector, the first resonances are the
vector mesons $\rho$ and $\omega$ at about $800$ MeV, $i.e.$ they are
considerably heavier than the Goldstone bosons. However, it is not only the
small octet--decuplet splitting which plays a role. One should also notice that
the $\Delta (1232)$ coupling to the $\pi N$ system is very large, $g_{\pi
\Delta N} \simeq 2 g_{\pi NN}$ with $g_{\pi NN} = 13.5$. Similarly, the
 $\gamma \Delta N$ coupling is very strong. Would the $\Delta(1232)$ (or the
decuplet) be weakly coupled to the nucleon (octet), its role would be very
different. It was therefore argued by Jenkins and Manohar [51]
to include the spin-3/2 decuplet in the effective theory
from the start. Denote by $T^\mu$ a Rarita-Schwinger fields in the heavy mass
formulation satisfying
$v\cdot T =0$. The effective Lagrangian of the spin-3/2
fields at lowest order reads
$${\cal L}_{MBT} = -i \bar T^{\mu} \, v\cdot {\cal D} \,T_\mu
+ \Delta  \, \bar T^{\mu } T_\mu + {C\over 2 } ( \bar T^{\mu }u_\mu  B + \bar
B u_\mu T^{\mu} )\,. \eqno(35)$$
where we have suppressed the flavor SU(3) indices.
For an explicit expression see ref.[65]. Notice that there is a
remaining mass dependence which comes from the average decuplet-octet splitting
$\Delta$ which does not vanish in the chiral limit. The constant $C$ is fixed
from the decay $\Delta \to N\pi$ or the average of some strong decuplet decays,
$|C| = 1.5 \ldots 1.9$.
The decuplet propagator carries the information about the mass splitting
$\Delta$ and reads
$${iP_{\mu \nu} \over v\cdot l - \Delta + i \epsilon}\eqno(36)$$
with
$P_{\mu\nu}$ a projector (for a review, see ref.[52]).
The appearance of the mass splitting $\Delta$ spoils the exact one--to--one
correspondence between the loop and low-energy expansion. The two
scales $F_\pi$ and
$\Delta$ which are both non-vanishing in the chiral limit enter the loop
calculations and they can combine in the form $(\Delta/F_\pi)^2$. The breakdown
of the consistent chiral counting in the presence of the decuplet is seen in
the loop contribution to the baryon mass. The loop diagrams with intermediate
decuplets states which naively count as order $E^4$ renormalize the average
octet baryon mass even in the chiral limit by an infinite amount. Therefore one
has to add a counter term of chiral power $E^0$ to keep the value $m_0$ fixed
$$\eqalign{\delta {\cal L}^{(0)}_{MB} & = - \delta m_0 \, \Tr (\bar B B)\cr
\delta m_0 & = {10\over 3} {C^2 \Delta^3 \over F_\pi^2} \biggl[ L + {1\over
16\pi^2} \bigl( \ln {2\Delta \over \lambda} - {5\over 6} \bigr) \biggr] \cr
L & = {\lambda^{d-4} \over 16 \pi^2} \biggl[ {1\over d-4} + {1\over 2} (
\gamma_E - \ln 4\pi - 1) \biggr] \cr } \eqno(37)$$
with $\lambda$ the scale introduced in dimensional regularization and $\gamma_E
$ the Euler-Mascheroni constant. This mass shift is similar to the
one in the relativistic version of pion-nucleon CHPT, where the non-vanishing
nucleon mass in the chiral limit leads to the same kind of complications [28].
 The inclusion  of the decuplet fields has
three effects on the mass formulae (31). First there is an infinite loop
contribution with decuplet intermediate states and, second,
an infinite renormalization of the order $E^2$ of the
low-energy constants $b_D, b_F$ and $b_0$ plus a finite contribution which
starts out at ${\cal O}(E^4)$. The constants $b_D$, $b_F$ and $b_0$ have
to be renormalized as follows:
$$\eqalign{
b_D  &= b_D^r(\lambda) - {\Delta C^2 \over 2F_p^2} L , \, \cr
b_F  &= b_F^r(\lambda) + {5\Delta C^2 \over 12F_p^2} L , \, \cr
b_0  &= b_0^r(\lambda) + {7\Delta C^2 \over 6F_p^2} L ,\cr}  \eqno(38)$$
where the finite pieces $b_{D,F,0}^r(\lambda)$ are then determined by the
fitting procedure. The explicit form of the decuplet contributions to the
octet masses can be found in refs.[44,45].
In table 1, I show some results
of these fits. It is obvious that simply taking the decuplet to account
for the $E^4$ (and higher) contributions does not lead to a consistent
picture of the scalar sector of CHPT (notice that in ref.[45] some tadpole
diagrams with insertions from ${\cal L}_{MB}^{(2)}$ have also been included but
that does not alter these conclusions). As already stressed a couple of times,
a {\it complete} $E^4$ calculation should be performed. For doing that, it
might be easier to use the decuplet to estimate some low--energy constants
rather than taking it as dynamical dof's in the EFT. The role of the decuplet
contributions has also been critically examined by Luty and White [53].
$$\hbox{\vbox{\offinterlineskip
\def\strut{\hbox{\vrule height 12pt depth 12pt width 0pt}}
\hrule
\halign{
\strut\vrule# \tabskip 0.1in &
\hfil#\hfil &
\hfil#\hfil &
\hfil#\hfil &
\hfil#\hfil &
\vrule# &
\hfil#\hfil &
\hfil#\hfil &
\hfil#\hfil &
\hfil#\hfil &
\vrule# \tabskip 0.0in
\cr
&  $D$ &  $F$ &  $\Delta$ & $C$ &&
$\sigma_{KN}^{(1)}(0)$ & $\sigma_{KN}^{(2)}(0)$ & $\Delta \sigma_{\pi N}$
 & $\Sigma^p_s$ & \cr
\noalign{\hrule}
& 0.75 & 0.50 & 293 & 1.8 && 38  & -23  & 14.3 & 513 & \cr
& 0.75 & 0.50 & 293 & 1.5 && 86  &  28  & 12.3 & 419 & \cr
& 0.75 & 0.50 & 231 & 1.8 && 67  &  10  & 15.1 & 455 & \cr
& 0.75 & 0.50 & 231 & 1.5 && 106 &  51  & 12.8 & 379 & \cr
& 0.56 & $2D/3$ & 293 & $2D$ && 244 & 150 & 6.8 & 132 & \cr
& 0.56 & $2D/3$ & 231 & $2D$ && 255 & 163 & 7.1 & 110 & \cr
\noalign{\hrule}}}}$$
\smallskip
{\noindent\narrower \it Table 1:\quad Results of
the calculation including the full
decuplet intermediate states. The values of $D,F,\Delta$ and $C$ are input.
All dimensionful numbers are in MeV. Empirically, $\Delta \sigma_{\pi N} =
15$ MeV and $\Sigma^p_s = m_s <p|\bar{s}s|p> = 130$ MeV [46].
\smallskip}
\goodbreak
\bigskip
\noindent{\bf 8. SPECTRAL DISTRIBUTION OF NUCLEON FORM FACTORS}
\bigskip
When discussing the forward Compton amplitude, I
mentioned that the corresponding branch point related to the one--pion
threshold has itself a $1/m$ expansion in the heavy mass formulation which
disturbs the analytical structure of the amplitude. To take a closer look at
this problem, let us consider the chiral expansion of
the so--called (isovector) Pauli form factor $F_2^V (t)$. It is defined by the
matrix--element of the isovector--vector quark current,
$$<p'|\bar q \gamma_\mu {\tau^a \over 2} q |p> = \bar{u}(p') \biggl[ \gamma_\mu
\, F_1^V (t) + {i \sigma_{\mu \nu} k^\nu \over 2 m} \, F^V_2 (t) \biggr]
{\tau^a \over 2} u(p)          \eqno(39)$$
with $k = p' - p$ and $t = k^2$. As first observed by  Frazer and Fulco [54]
and discussed in detail by H\"ohler and Pietarinen
[55] the imaginary part of $F_2^V (t)$  exhibits a strong enhancement very
close to threshold ($t = 4 M_\pi^2$)
as shown in fig.6a. The imaginary part of the isovector
nucleon form factors inherit the singularity  on the second sheet due to the
projection of the  Born term (at $t_0 = 4 M_\pi^2 ( 1 - M_\pi^2 / 4 m^2) =
3.98 M_\pi^2$) in $\pi N$ scattering
(from diagrams of the type $\gamma \to \pi \pi \to \bar{N} N$).

Let us first consider the chiral expansion of Im~$F_2^V (t)$ in the
relativistic formulation of baryon CHPT. Following Gasser et al. [28], one has
$${\rm Im} F_2^V (t) = {8 g_A^2 \over F_\pi^2} m^4 \biggl[ 4 \,
{\rm Im} \, \gamma_4
(t) + {\rm Im} \, \Gamma_4 (t) \biggr]             \eqno(40)$$
where the loop functions and their imaginary parts
$\gamma_4$ and $\Gamma_4$ are given in ref.[28]. For our purpose, we only need
Im~$\gamma_4 (t)$ since its threshold is the two--pion cut whereas Im~$\Gamma_4
(t)$ only starts to contribute at $t = 4 m^2$. The resulting imaginary part
for Im~$F_2^V (t) / t^2$ is shown in fig.6b (solid line). One sees that the
strong increase at threshold is reproduced (see also the remarks in ref.[28])
since the chiral representation of Im $\gamma_4 (t)$ indeed has the proper
analytical structure, i.e. the singularity on the second sheet at $t_0$.
The chiral representation of Im~$F_2^V (t)/ t^2$ does not stay constant on the
left shoulder of the $\rho$--resonance but rather drops.
This is due to the fact that
in the one loop approximation, one is only sensitive to the first  term in
the chiral expansion of the pion charge form factor $F_\pi^V (t)$.
In fig.6a I also show calculations with $F_\pi^V (t) = 1$ (dashed line) and
$F_\pi^V (t) = 1 + <r_\pi^2> \, t /6$ (dash--dotted line). These curves
resemble
very much the chiral expansion. To reiterate, this particular example shows
that in the relativistic version of baryon CHPT the pertinent analytical
structures of current and S--matrix elements  are given correctly.
\midinsert
\vskip 17.0truecm
{\noindent\narrower \it Fig.~6:\quad
(a) Dispersion--theoretical result for Im~$F_2^V (t)/ t^2$ ($t$ in units of the
pion mass, denoted $\mu$ here). As  D--function, the inverse pion form factor
is used, $D(t) = 1/F_\pi^V (t)$. The dash--dotted and dashed lines are
explained in the text. (b) Chiral representation in the relativistic
formulation of baryon CHPT [28] (solid line) and in the heavy mass approach
(dashed line) [31].
\smallskip}
\vskip -0.5truecm
\endinsert
Let us now consider the heavy mass approach. The corresponding imaginary part
follows from ref.[31],
$${\rm Im} F_2^V (t) = {g_A^2 m \over 8 F_\pi^2} \biggl( {1 \over 4} - {M_\pi^2
\over t}\biggr) \sqrt{t}     \eqno(41)$$
Here, the imaginary part comes form a $\ln (2M_\pi - \sqrt{t})$ which has a
branch point at $t = 4 M_\pi^2$ (chiefly because to lowest order in the $1/m$
expansion the threshold energy of $\pi N$ scattering is $\omega_0 = M_\pi$
 [39] and the corresponding left--handed cut starts there).
This also leads to an enhancement of the
imaginary part of $F_2^V (t)$ as shown by the dashed line in fig.6b.
The enhancement is stronger than in the relativistic case.
 Stated differently,
to this order in the chiral expansion the analytic structure is not given
correctly much like in the case of the forward Compton scattering amplitude
discussed before. One should therefore perform an order $E^5$ calculation in
the heavy mass approach to have a sufficently accurate and {\it correct}
representation of the isovector nucleon form factors. A two loop calculation
will also answer the yet unresolved question whether or not in the isoscalar
channel there is an enhancement around $t = 9 M_\pi^2$. State of the art
dispersion theoretical analysis of the nucleon form factors assume only a set
of poles in the corresponding spectral distributions [56]. Finally, I wish
to stress that in this context
the matching formalism discussed in ref.[31] starts to play a role (which
allows to relate matrix--elemnents in the heavy mass and relativistic
formulation of CHPT).
\bigskip
\noindent{\bf 9. ASPECTS OF ELECTROWEAK PION PRODUCTION}
\bigskip
In this section, I will discuss some physics aspects related to threshold pion
production by electroweak interactions. This is of particular interest for the
now existing generation of CW electron machines like at Mainz, Bates $\ldots$.

Let me first consider the production of one single pion by the isovector axial
current. As particularly stressed by Adler [57], a unified treatment of
em and weak pion production allows to relate information from neutrino--nucleon
and electron--nucleon  scattering experiments. Obviously, by using PCAC, the
coupling of the weak axial current to the nucleon in the initial state and a
nucleon plus a pion in the final state is closely related to $\pi N$
scattering. Consider now a reaction of the type
$\nu (k_1) + N(p_1) \to l (k_2) + N(p_2) + \pi^a (q)$  and define $k = k_1 -
k_2$. The threshold energy squared is $s = (p_1 +k)^2 =(m + M_\pi)^2$.
 At threshold and in the $\pi N$ cms frame, one can express the pertinent
matrix--element in terms of six S--wave multipoles,
$$T^{(\pm)}\cdot \epsilon =
 4 \pi ( 1 + \mu ) \chi_2^\dagger \bigl[
\epsilon_0 L_{0+}^{(\pm )} + \epsilon \cdot k H_{0+}^{(\pm )} + i \vec{\sigma}
\cdot (\hat{k} \times \vec{\epsilon} ) M_{0+}^{(\pm )} \bigr] \chi_1
\eqno(42)$$
with $\chi_{1,2}$ two--component Pauli  spinors, $\mu = M_\pi / m$
the ratio of the pion and nucleon mass and $\epsilon_\mu \sim \bar{u}_l
\gamma_5 \gamma_\mu u_\nu$ the axial polarization vector.
The superscript '$\pm$' refers to the isospin even/odd part of the amplitude.
In ref.[58], we
derived the chiral expansions of these threshold multipoles to order $E^3$.
Of particular interest is the multipole $L_{0+}^{(+)}$ since it is directly
proportional to the so--called scalar form factor of the nucleon,
 $\sigma_{\pi N}
(t)$ $ \sim$ $ <p'|\hat{m}(\bar u u + \bar d d)|p>$ [58],
$$L_{0+}^{(+)} = {1 \over 3 \pi M_\pi F_\pi} \biggl\{ \sigma_{\pi N}
(k^2 - M_\pi^2) -
{1 \over 4} \sigma_{\pi N} (0) \biggr\} - {a^+ F_\pi \over M_\pi} - {g_A^2
M_\pi \over 16 \pi m F_\pi} + C^{(+)}_L M_\pi^2 + {\cal O}(q^3) \eqno(43)$$
The constant $C^{(+)}$ subsumes numerous $k^2$--independent
kinematical, loop and counter term  contributions. If one assumes $C^{(+)}$ to
be of the order of 1 GeV$^{-3}$, the term proportional to the scalar form
factor dominates the amplitude (43) in the threshold region. This might offer
another determination of this much discussed quantity. However, an analysis
including also higher order effects has to be performed to find out how
cleanly this multipole can be separated in
neutrino--induced pion production and how
large the corresponding cross section is. It is furthermore
interesting to note that
although $L_{0+}^{(+)}$ vanishes at the photon point $k^2 = 0$ in the chiral
limit, its slope nevertheless stays finite - this is a particular effect due to
chiral loops.

Another reaction of interest is the  photoproduction of two pions in the
threshold region. It gives complimentary information to the extensive
studies of single pion photo-- and electroproduction performed over the last
few years. Dahm and Drechsel [59] were the first to systematically study the
process $\gamma N \to \pi \pi N$ in a chiral field theory. To be specific, they
considered Weinberg's pion--nucleon Lagrangian [60] coupling in the photon via
minimal substitution. At threshold, the transition current takes the form
$$ T \cdot \epsilon \bigl|_{\rm thr} =
 i \svec{\sigma} \cdot (\svec{\epsilon} \times
\svec{k}) \, [ M_1 \delta_{ab}
 + M_2 \delta_{ab} \tau_3 + M_3 (\tau_a \delta_{b3}
+ \tau_b \delta_{a3} )]    \eqno(44) $$
where 'a,b' are the pion isospin indices and the $\tau$'s act
on the nucleon. The
explicit form of the corresponding five--fold differential cross section is
given in [59]. Here, we are interested in the chiral expansion of the
multipoles $M_1$, $M_2$ and $M_3$, i.e. their expansion in powers of $M_\pi$.
In general, the reaction $\gamma N \to \pi \pi N$ involves five independent
Mandelstam variables. At threshold, the kinematics is simplified since the
pion four--momenta are equal, $q_1 = q_2 = (M_\pi,0,0,0)$. In heavy fermion
CHPT the calculation furthermore simplifies if one works in the Coulomb gauge
$\epsilon \cdot v = 0$ and realizes that $S \cdot q_1 = S \cdot q_2 = 0$ [61].
The lowest order result stems from tree diagrams with one insertion from ${\cal
L}_{\pi N}^{(2)}$ as shown in fig.7.
\midinsert
\vskip 7.0truecm
{\noindent\narrower \it Fig.~7:\quad
Tree diagrams which lead to eq.(45). The box denotes an insertion from ${\cal
L}_{\pi N}^{(2)}$. Solid, dashed and wiggly lines denote nucleons, pions and
photons, respectively.
\smallskip}
\vskip -0.5truecm
\endinsert
One finds
$$ M_1 = {\cal O}(M_\pi) , \, M_2 = {e \over 4 m F_\pi^2} (2 g_A^2 - 1
-\kappa_V) + {\cal O}(M_\pi), \, M_3 = -{M_2 \over 2}  \, \, .   \eqno(45)$$
These differ from the results in [59] by the terms proportional to $\kappa_V$
and are a factor two smaller in magnitude. The reason is that gauging the
Weinberg Lagrangian can not generate the anomalous couplings of the photon.
Of course, simply calculating these tree diagrams is not sufficient, one has to
at least work out the ${\cal O}(M_\pi)$ corrections. These are $(i)$
kinematical corrections of the type $M_\pi / m$, $(ii)$ contributions from
one--loop diagrams, $(iii)$ further insertions from ${\cal L}_{\pi N}^{(2,3)}$
and $(iv)$ contributions from tree diagrams with intermediate $\Delta(1232)$
states. As a preliminary result, let me consider the corrections of the first
two types (a more thorough discussion can be found in ref.[61]). Although the
$\Delta(1232)$ is very close to the two pion production threshold (the energy
difference being 17 MeV), the potentially large diagrams with small energy
denominators of the type $m_\Delta - (m_N + 2M_\pi )$ are suppressed by
corresponding numerators. The corrections of type $(i)$ and $(ii)$ lead to [61]
$$\eqalign{ M_1 &= {e g_A^2 M_\pi \over 4 m^2 F_\pi^2} = 0.019 \, {\rm fm}^3
\cr
M_2 &= {e M_\pi \over 4 m^2 F^2_\pi}(g_A^2 - \kappa_V) +
{e g_A^2 M_\pi \over 64 \pi  F^4_\pi}\biggl[{3 \pi \over 2} + i( 2 \sqrt{3} -
\ln (2+\sqrt{3}))\biggr] \cr &= (0.159 + i \, 0.082) \, {\rm fm}^3 \cr
M_3 &= {e M_\pi \kappa_V \over 8 m^2 F^2_\pi} +
{e g_A^2 M_\pi \over 256 \pi  F^4_\pi}\biggl[6 -{3 \pi \over 2}
+ i( 2 \sqrt{3} -
\ln (2+\sqrt{3}))\biggr] = (0.032 + i \, 0.020) \, {\rm fm}^3 \cr} \eqno(46)$$
For $M_1$, the loops do not contribute to order $M_\pi$ where as for $M_{2,3}$
they lead to a complex correction. This due to some loop diagrams involving two
or three  pion propagators which acquire an imaginary part for $\omega > M_\pi$
(here, $\omega = 2 M_\pi$). The first term on the r.h.s. of (46) stems from the
kinematical corrections. Comparing the numbers in (46) to the lowest order
results $M_1 = 0$, $M_2 = - 2 M_3 = 0.084$ fm$^3$, one sees that the $M_\pi$
corrections are large. One might therefore question the whole approach, but it
is conceivable that once the type $(iv)$ corrections from the $\Delta$ are
included, the dominant physics will be under control and subsequent higher
order corrections play a minor role. This topic is under investigation [61].
To get an idea about the role of the loop and kinematical corrections, let
us consider the specific final states like in $\gamma p \to \pi^0 \pi^0 p$.
In that case, the cross section is proportional to the quantity $|{\cal M}_{00}
| = [(M_1+ M_2+2M_3)(M_1 + M_2+2M_3)^*]^{1/2}$
 and similarly for the other channels. In
table 2, I show $|{\cal M}_{ij}|$ in fm$^3$ for the lowest order (45) and
with the $M_\pi$ corrections (46). The most prominent result is that the double
$\pi^0$ channels, which are vanishing to lowest order, are in fact  dominant
$$\hbox{\vbox{\offinterlineskip
\def\strut{\hbox{\vrule height 12pt depth 12pt width 0pt}}
\hrule
\halign{
\strut\vrule# \tabskip 0.1in &
\hfil#\hfil &
\vrule# &
\hfil#\hfil &
\hfil#\hfil &
\vrule# \tabskip 0.0in
\cr
&  process &&  ${\cal O}(1)$ & ${\cal O}(1) + {\cal O}(M_\pi)$ & \cr
\noalign{\hrule}
& $\gamma p \to \pi^+ \pi^- p$ && 0.084 & 0.196 & \cr
& $\gamma p \to \pi^+ \pi^0 n$ && 0.059 & 0.053 & \cr
& $\gamma p \to \pi^0 \pi^0 p$ && 0.000 & 0.262 & \cr
& $\gamma n \to \pi^+ \pi^- n$ && 0.084 & 0.162 & \cr
& $\gamma n \to \pi^- \pi^0 p$ && 0.059 & 0.053 & \cr
& $\gamma n \to \pi^0 \pi^0 n$ && 0.000 & 0.229 & \cr
\noalign{\hrule}}}}$$
\smallskip
{\noindent\narrower \it Table 2:\quad Contribution of the threshold
multipoles to two pion production channels in fm$^3$ (no phase space
factors are accounted for).
\smallskip}
\noindent after the inclusion of
the ${\cal O}(M_\pi)$ corrections. This is completely
different from the single photoproduction case in which the final states
including a charged pion have the largest cross sections. An experimental
verification of this pattern would be of utmost interest. It also persists
when one includes the $\Delta$--corrections at this order in the chiral
expansion [61]. The corresponding total cross sections for the various channels
can be compactly written as [61]
$$\sigma_{\rm tot}^{\gamma N \to \pi \pi N} (E_\gamma ) = {M_\pi^2 (1+ \mu )
\over 32 \pi^2 (1+ 2 \mu )^{11/2}} (E_\gamma - E_\gamma^{\rm thr})^2
|\eta_1 M_1 + \eta_2 M_2 + \eta_3 M_3 |^2 \xi               \eqno(47)$$
with $E_\gamma$ the photon energy in the lab frame, $\eta_{1,2,3}$ isospin
factors (like e.g. $\eta_1 = \eta_2 =1 , \eta_3 =2$ for the $p \pi^0 \pi^0$
final state) and $\xi$ is a Bose factor ( = 1/2 in case of equal particles in
the final state, one otherwise). The threshold energy is $E_\gamma^{\rm thr} =
2 M_\pi ( 1 + \mu ) = 320.7$ MeV with $\mu = M_\pi /m$. The formula (47) is
only
valid close to threshold  assuming that the amplitude in the threshold region
can be approximated by the threshold amplitude. Furthermore, the three--body
phase space has been approximated by an analytical expression which is good
within a few percent. For $E_\gamma = 330$ MeV, i.e. 10 MeV above threshold,
the total cross section is 0.33, 0.37 and 0.03 nbarn for $\gamma p \to p \pi^0
\pi^0, p \pi^+ \pi^-$ and $n \pi^+ \pi^0$, in order. Of course, these numbers
should only be taken as a first approximation due to the assumptions going
into their calculation.   It is therefore a tough
experimental task to measure these reactions close to threshold and verify the
effects of the chiral loops.
\bigskip \vfill \eject
\noindent{\bf 10. BRIEF OUTLOOK}
\bigskip
In this lecture I could only give a glimpse of the many facets of chiral
perturbation theory. The role of effective Lagrangian methods in the
parametrization of physics beyond the standard model and in the context of
longitudinal vector boson scattering to test the Higgs sector of the
electroweak symmetry breaking has been discussed here by  Burgess [66] and
Phillips [67]. Another widely discussed topic is the combined application of
chiral symmetry and heavy quark effective field theory methods (some references
can be traced back from [4]). Furthermore, these methods allow also to make
precise statements for finite temperature and volume effects and much more.
There remain many open theoretical problems and challenging experimental tasks
to further tighten our understanding of the strong interactions at momentum
scales were they are really {\it strong}.
\bigskip
\bigskip \bigskip
\noindent I would like to thank the organizers, in particular Dr. Uma Shankar,
 for the warm hospitality
extended to me and an efficient organization. I also thank
the Saha Institute for Nuclear Physics, in particular Prof. S. Mallik, for
providing a stimulating atmosphere and support to perform part of this work.
\bigskip \bigskip \bigskip
\noindent{\bf REFERENCES}
\medskip
\item{1.}S. Weinberg, {\it Phys. Rev. Lett.\/} {\bf 17} (1966) 616.
\smallskip
\item{2.}S. Weinberg, {\it Physica} {\bf 96A} (1979) 327.
\smallskip
\item{3.}J. Gasser and H. Leutwyler, {\it Ann. Phys. {\rm (N.Y.)}\/}
 {\bf 158} (1984) 142;  {\it Nucl. Phys.\/} {\bf B250} (1985) 465.
\smallskip
\item{4.}Ulf-G. Mei{\ss}ner, {\it Rep. Prog. Phys.\/} {\bf 56} (1993) 903.
\smallskip
\item{5.}H. Leutwyler, Bern University preprint BUTP-93/24, 1993. \smallskip
\item{6.}J. Gasser and H. Leutwyler,
{\it Phys. Lett.\/} {\bf 125B} (1983) 325.
\smallskip
\item{7.}W. Ochs, Max--Planck--Institute preprint MPI--Ph/Ph 91--35, 1991.
\smallskip
\item{8.}N.  H. Fuchs, H. Szadijan and J. Stern, {\it Phys. Lett.\/} {\bf
B269} (1991) 183. \smallskip
\item{9.}N.  H. Fuchs, H. Szadijan and J. Stern, {\it Phys. Rev.\/} {\bf
D47} (1993) 3814. \smallskip
\item{10.}L. L. Nemenov, {\it Yad. Fiz.\/} {\bf 41} (1985) 980;

J. Uretsky and J. Palfrey, {\it Phys. Rev.\/} {\bf 121} (1961) 1798;

G. Czapek et al., letter of intent CERN/SPSLC 92--44, 1992.
\smallskip
\item{11.}R. Altmeyer et al., J\"ulich preprint HLRZ 92--17, 1992.
\smallskip
\item{12.}J. Gasser and Ulf-G. Mei{\ss}ner, {\it Phys. Lett.\/} {\bf B258}
(1991) 219.
\smallskip
\item{13.}V. Bernard, N. Kaiser and Ulf-G. Mei{\ss}ner,
{\it Nucl. Phys.\/} {\bf B357} (1991) 129;
{\it Phys. Rev.\/} {\bf D43} (1991) R3557.
\smallskip
\item{14.}A. Dobado and J. R. Pelaez,
{\it Phys. Rev.\/} {\bf D47} (1993) 4883. \smallskip
\item{15.}J. F. Donoghue, J. Gasser and H. Leutwyler, {\it Nucl. Phys.\/}
{\bf B343} (1990) 341.
\smallskip
\item{16.}J. Gasser and Ulf-G. Mei{\ss}ner,
{\it Nucl.Phys.\/} {\bf B357} (1991) 90.
\smallskip
\item{17.}J. Bijnens and F. Cornet, {\it Nucl. Phys.\/} {\bf B296} (1988) 557.
\smallskip
\item{18.}J. F. Donoghue, B. R. Holstein and Y. C. R. Lin,  {\it Phys. Rev.
\/} {\bf D37} (1988) 2423.
\smallskip
\item{19.}H. Marsiske et al.,  {\it Phys. Rev.\/} {\bf D41} (1990) 3324.
\smallskip
\item{20.}M.R. Pennington, in The DA$\Phi$NE Physics Handbook, eds. L. Maiani,
G. Pancheri and N. Paver, Frascati, 1992. \smallskip
\item{21.}J. F. Donoghue and B. R. Holstein,
 {\it Phys. Rev.\/} {\bf D48} (1993) 37. \smallskip
\item{22.}S. Belluci, J. Gasser and M.E. Sainio, Bern University preprint
BUTP-93/18, 1993. \smallskip
\item{23.}G. Ecker, A. Pich and E. de Rafael, {\it Phys. Lett.\/} {\bf
B189} (1987) 363; {\it Nucl. Phys.\/} {\bf B291} (1987) 692;
{\it Nucl. Phys.\/} {\bf B303} (1988) 665.
\smallskip
\item{24.}G.D. Barr et al.,  {\it Phys. Lett.\/} {\bf B242} (1990) 523;
{\bf B284} (1992) 440;

V. Papadimitriou et al., {\it Phys. Rev.\/} {\bf D44} (1991) 573.
\smallskip
\item{25.}A. Cohen, G. Ecker and A. Pich,  {\it Phys. Lett.\/}
{\bf B304} (1993) 347.  \smallskip
\item{26.}J. Kambor and B.R. Holstein, Amherst preprint UMHEP--397, 1993.
\smallskip
\item{27.}
Ulf-G. Mei{\ss}ner, {\it Int. J. Mod. Phys.} {\bf E1} (1992) 561.
\smallskip
\item{28.}J. Gasser, M.E. Sainio and A. ${\check {\rm S}}$varc,
{\it Nucl. Phys.\/} {\bf B307} (1988) 779.
\smallskip
\item{29.}A. Krause, {\it Helv. Phys. Acta\/} {\bf 63} (1990) 3.
\smallskip
\item{30.}E. Jenkins and A.V. Manohar, {\it Phys. Lett.\/} {\bf B255} (1991)
558. \smallskip
\item{31.}V. Bernard, N. Kaiser, J. Kambor
and Ulf-G. Mei{\ss}ner, {\it Nucl. Phys.\/} {\bf B388} (1992) 315.
\smallskip
\item{32.}G. Ecker, {\it Czech. J. Phys.}, in print. \smallskip
\item{33.}G. Ecker, private communication. \smallskip
\item{34.}M. Damashek and F. Gilman, {\it Phys. Rev.\/} {\bf D1} (1970) 1319;

V.A. Petrunkin, {\it Sov. J. Nucl. Phys.\/} {\bf 12} (1981) 278;

T.A. Armstrong et al., {\it Phys. Rev.\/} {\bf D5} (1970) 1640.
\smallskip
\item{35.}A. Zieger {\it et al.}, {\it Phys. Lett.\/} {\bf B278}
(1992) 34;

F.J. Federspiel {\it et al.}, {\it Phys. Rev. Lett.\/} {\bf 67}
(1991) 1511;

E.L. Hallin {\it et al.}, {\it Phys. Rev.\/} {\bf C48} (1993) 1497.

K.W. Rose {\it et al.}, {\it Phys. Lett.\/} {\bf B234}
(1990) 460;

J. Schmiedmayer {\it et al.}, {\it Phys. Rev. Lett.\/} {\bf 66}
(1991) 1015.
\smallskip
\item{36.}V. Bernard, N. Kaiser and Ulf-G. Mei{\ss}ner,
{\it Phys. Rev. Lett.\/} {\bf 67} (1991) 1515; {\it Nucl. Phys.\/}
{\bf B373} (1992) 364. \smallskip
\item{37.}J. Gasser and A. Zepeda, {\it Nucl. Phys.\/} {\bf B174}
(1980) 445. \smallskip
\item{38.}V. Bernard, N. Kaiser, A. Schmidt
and Ulf-G. Mei{\ss}ner, {\it Phys. Lett.\/} {\bf B319} (1993) 261.
\smallskip
\item{39.}V. Bernard, N. Kaiser and Ulf-G. Mei{\ss}ner,
{\it Phys. Lett.\/} {\bf B309} (1993) 421. \smallskip
\item{40.}G. Ecker, J. Gasser, A. Pich and E. de Rafael,
{\it Nucl. Phys.\/} {\bf B321} (1989) 311.
\smallskip
\item{41.}V. Bernard, N. Kaiser, A. Schmidt
and Ulf-G. Mei{\ss}ner, preprint CRN-93/55, 1993. \smallskip
\item{42.}V. Bernard, N. Kaiser, J. Kambor and Ulf-G. Mei{\ss}ner,
{\it Phys. Rev.\/} {\bf D46} (1992) 2756. \smallskip
\item{43.}H.J. Lipkin and M. Moinester, {\it Phys. Lett.\/} {\bf B287}
(1992) 179.   \smallskip
\item{44.}V. Bernard, N. Kaiser and Ulf-G. Mei{\ss}ner,
{\it Z. Phys.\/} {\bf C60} (1993) 111.  \smallskip
\item{45.}E. Jenkins, {\it Nucl. Phys.\/} {\bf B368} (1992) 190.
\smallskip
\item{46.}J. Gasser, H. Leutwyler and M.E. Sainio, {\it Phys. Lett.\/}
 {\bf 253B} (1991) 252, 260. \smallskip
\item{47.} J.N. Labrenz and S.R. Sharpe, University of Washington preprint
UW/PT-93-07, 1993. \smallskip
\item{48.}R.L. Jaffe and C. Korpa, {\it Comments Nucl. Part. Phys.\/} {\bf 17}
(1987) 163. \smallskip
\item{49.}P. Gensini, {\it $\pi N$ Newsletter} {\bf 6} (1992) 21.
\smallskip
\item{50.}R. Dashen, E. Jenkins and A.V. Manohar, preprint UCSD/PTH 93-21,
1993. \smallskip
\item{51.}E. Jenkins and A.V. Manohar, {\it Phys. Lett.\/} {\bf B259} (1991)
353. \smallskip
\item{52.}E. Jenkins and A.V. Manohar, in "Effective field theories of the
standard model", ed. Ulf--G. Mei{\ss}ner, World Scientific, Singapore,
1992. \smallskip
\item{53.}M. Luty and A. White, {\it Phys. Lett.\/} {\bf B319} (1993)
261. \smallskip
\item{54.}W.R. Frazer and J. Fulco, {\it Phys. Rev.\/}
{\bf 117} (1960) 1603, 1609. \smallskip
\item{55.}G. H\"ohler and E. Pietarinen, {\it Phys. Lett.\/} {\bf B53} (1975)
471. \smallskip
\item{56.}G. H\"ohler, in Landolt--B\"ornstein, vol.9 b2, ed. H. Schopper
(Springer, Berlin, 1983). \smallskip
\item{57.}S.L. Adler, {\it Ann.Phys. {\rm (N.Y.)}\/} {\bf 50} (1968) 189.
\smallskip
\item{58.}V. Bernard, N. Kaiser and Ulf--G. Mei{\ss}ner, CRN preprint
CRN--93/56, 1993. \smallskip
\item{59.}R. Dahm and D. Drechsel, in Proc. Seventh Amsterdam Mini--Conference,
eds. H.P. Blok, J.H. Koch and H. De Vries, Amsterdam, 1991. \smallskip
\item{60.}S. Weinberg, {\it Phys. Rev.\/} {\bf 166} (1968) 1568.
\smallskip
\item{61.}V. Bernard, N. Kaiser, Ulf--G. Mei{\ss}ner and A. Schmidt,
in preparation. \smallskip
\item{62.}V. Bernard, N. Kaiser and Ulf--G. Mei{\ss}ner, {\it Nucl. Phys.\/}
{\bf B364} (1991) 283. \smallskip
\item{63.}F. Butler et al., {\it Phys. Rev. Lett.\/} {\bf 70} (1993) 2849.
\smallskip
\item{64.}M. Golterman, these proceedings. \smallskip
\item{65.}V. Bernard, N. Kaiser and Ulf--G. Mei{\ss}ner, {\it Phys. Rev.\/}
{\bf D48} (1993) 3062. \smallskip
\item{66.}C. Burgess, these proceedings. \smallskip
\item{67.}R.J.N. Phillips, these proceedings. \smallskip

\vfill \eject \end